\documentclass[12pt]{article}

\usepackage{indentfirst}
\usepackage{amsmath}
\usepackage{amssymb}
\usepackage{amsfonts}
\usepackage{amscd}
\usepackage{amsbsy}
\usepackage{amsthm}
\usepackage{latexsym}
\usepackage{graphicx,color} 
\usepackage[dvipsnames]{xcolor}
\usepackage[colorlinks]{hyperref}
\hypersetup{linkcolor=blue,citecolor=blue,urlcolor=blue}
\def\nn{\nonumber}       
\def\beq{\begin{eqnarray}}
\def\eeq{\end{eqnarray}}
\def\ln{\,\mbox{ln}\,}

\def\Det{\,\mbox{Det}\,}

\def\Tr{\,\mbox{Tr}\,}

\DeclareMathOperator{\cx}{\square}

\def\al{\alpha}
\def\be{\beta}

\def\vp{\varepsilon}
\def\ep{\epsilon}

\def\la{\lambda}

\def\pa{\partial}

\def\si{\sigma}

\def\ph{\varphi}

\def\th{\theta}

\def\Ga{\Gamma}

\def\La{\Lambda}

\usepackage[symbol]{footmisc} 
\usepackage{float} 
\usepackage{cite}  
\usepackage{titlesec}

\usepackage{ulem} 

\titleformat*{\section}{\large\bfseries}
\titleformat*{\subsection}{\normalsize\bfseries}

\begin{document}

\begin{center}
\renewcommand*{\thefootnote}{\fnsymbol{footnote}}
{\LARGE 
Derivative expansion in a two-scalar field theory}
\vskip 6mm

{\bf Al\'{\i}cia G. Borges}\footnote{E-mail address:
\ aliciagomes1a1@gmail.com}
\quad
and
\quad
{\bf Ilya L. Shapiro}\footnote{E-mail address: \ ilyashapiro2003@ufjf.br}

\vskip 5mm

{\sl Departamento de F\'{\i}sica, ICE,
Universidade Federal de Juiz de Fora}

{\sl Campus Universit\'{a}rio - Juiz de Fora, 36036-900, MG, Brazil}
\vskip 8mm

\textbf{Abstract}

\end{center}

\begin{quotation}
\noindent
The derivative expansion of the effective action is considered in the
model with two interacting real scalar fields in curved spacetime.
Using the functional approach and local momentum representation, the
coefficient of the derivative term is calculated up to the first order in
curvature in the one-scalar theory. The two-scalar problem is solved
by extracting normal modes and consequent reduction to the
single-scalar case. The method can be applied to a larger number of
scalars. In the theory with strong hierarchy of masses, the
renormalized effective potential and the coefficients of the
second-order derivative terms demonstrate the quantum decoupling
in the low-energy limit.
\vskip 4mm

\vskip 2mm


\end{quotation}

\vskip 2mm	

\setcounter{footnote}{0}
\renewcommand*{\thefootnote}{\arabic{footnote}}

\section{Introduction}
\label{sec1}

Certain cosmological models, such as assisted inflation, are based
on two scalar fields with a strong hierarchy of masses (see, e.g.,
Refs.~\cite{Liddle_1998,Wands,Peterson:2010np}
and the more recent reference \cite{Heatal}). Since the typical
energy scale in the inflationary epoch is high, it may be important to
evaluate quantum effects modifying the scalar action. The relevant
quantum corrections may include not only the effective potential of
scalars but also the lowest-order derivative terms. In this respect,
one can note that there are several insufficiently studied aspects of
the quantum theory. The first one of these important details is that
cosmological inflation is characterized by large values of
curvatures, which cannot be neglected in the effective potential
and in the quantum contributions to the derivative terms. The
curvature-dependent terms are pretty well-known for the effective
potential, but almost not explored for the derivative expansion. The
last unexplored (at least to our knowledge) issue is that the quantum
effects may be quite different in the high-energy (UV) and
low-energy (IR) limits owing to the quantum decoupling of the
large-mass scalar in the latest stage of inflation, where the energy
scale gets lower.  As a result, at the later phase of inflation, the
quantum effects of only one scalar remain relevant.
The described features and possibilities make it interesting to
explore the curvature-dependent effective potential and the
lowest-order derivative expansion of the effective action in the
multi-scalar models, especially with different masses.

The derivative expansion of the effective action is a traditional
subject of study in quantum field theory
\cite{iliopoulos1975,fraser1985},
which has been extensively applied in various branches of effective
field theory (see, e.g., \cite{Wein-II} and \cite{Manohar-2000}). The
technique was also developed in curved spacetime using Riemann
normal coordinates \cite{Hu-1984} and zeta-regularization
\cite{Kirsten-1993}. It is also worth noting Ref.~\cite{Cheyette-1985}
and \cite{Chan} with the development of the derivative expansion
for $O(N)$ scalar model  in flat spacetime, i.e., $N$ copies of a scalar
field with equal masses.

In the present work, we explore the one-loop quantum contributions
to the effective action of two scalar fields with different masses in
zero and second orders of derivative expansion, in curved spacetime.
We follow a simple approach similar to extracting normal modes of
the oscillator in classical mechanics. The scheme can be applied to
the multiple scalars and arbitrary self-interactions, at least without
derivatives. Following the previous
literature \cite{Hu-1984}, we use functional methods  in curved
spacetime, combined with local momentum representation
\cite{Bunch1979} (see also \cite{OUP} for further references), based on
normal coordinates \cite{Petrov} (see also \cite{Tensors} for a more
pedagogical introduction) to get the result in an arbitrary curved
metric background.

One of the motivations for our work is to elaborate the formalism
for taking into account quantum effects in the cosmological models
with multiple scalar fields. Such quantum corrections may influence
inflationary dynamics and therefore may be relevant. An interesting
aspect in these scenarios is the potential (IR) decoupling of heavy
fields, which, if present, could effectively reduce the number of
relevant degrees of freedom at late times. This phenomenon is
particularly important for understanding whether in the later
cosmological epoch, only the quantum effects of a lighter scalar
field remain significant, reinforcing the role of effective quantum
approach in such kind of models. Beyond the applications to
inflation, our results may be also useful for a broader understanding
of quantum corrections in multi-field theories, where functional
techniques provide a powerful framework for systematically extracting
both potential and lowest-order derivative terms.

The paper is organized as follows.
In Sec.~\ref{sec2} we follow the work \cite{fraser1985} on the
derivative expansion for a single scalar field. After this, the result
of this consideration is extended to an arbitrary curved spacetime
up to the first order in curvature, by using normal coordinates.
Sec.~\ref{sec3} reports on the derivative expansion in the model
with two scalars with different masses. We perform the calculations
in a direct way and also follow the new ``double diagonalization''
approach to reduce the problem to a single scalar case. It is
remarkable that this method is more economical and, in principle,
permits a generalization to an arbitrary number of scalars.
Sec.~\ref{sec4} is devoted to the decoupling of massive degree
of freedom in the theory with a strong hierarchy of masses.
Finally, in Sect.~\ref{Conc} we draw our conclusions.
Throughout this work, we adopt the metric signature(+,-,-,-)
and assume the Wick rotation where necessary, in some cases,
without additional comments.

\section{ Derivative expansion for a single scalar field }
\label{sec2}

Derivative expansion is a useful extension of the functional
methods to calculate the effective potential. In general, the
effective action is a nonlocal functional, but one can start from
the constant part, i.e., the effective potential, and extract the
lowest-order terms corresponding to the weakly varying fields.
The scheme may be applied to different kinds of fields, but the
most common application is to scalars. In the case of a single
mean field $\ph$, the first terms in the expansion have the form
\beq
\Ga[\ph,\,g_{\mu\nu}]\,=\,
\int d^4x\,\sqrt{-g}\,\,\Big\{
- V_{eff}(\phi) \,+\, \frac12\,Z(\phi)
\,g^{\mu\nu}\,\pa_\mu\phi\,\pa_\nu\phi
+ ... \,\Big\}\,.
\label{EffPot}
\eeq
The calculation of $\,V_{eff}(\phi)\,$ can be performed for a
constant $\phi=\phi_0$ to any order of the loop expansion and
for the theories with different contents of quantum fields. In this
work, we restrict our attention to the self-interacting scalar field
in curved space-time. The one-loop flat-space result was derived
in the renowned work \cite{ColeWein}. The general investigation
based on the functional methods, was done in \cite{Jackiw74}.
Let us reproduce the expression for the renormalized one-loop
contribution to the effective potential derived in \cite{CorPot}
by using the mass-dependent scheme of renormalization and
normal coordinates,
\beq
V_{eff}^{ren}(g_{\mu\nu},\,\ph)
&=& \rho_\La \,+\,\frac12\,(m^2-\xi R)\ph^2 \,+\, V
\label{REN-potya}
\\
&+&
\frac{1}{2(4\pi)^2}\,
\Big[\frac12\,\big(m^2 + V^{\prime\prime} \big)^2
\,-\, \Big(\xi-\frac16\Big)\, R\,
\big(m^2+V^{\prime\prime}\big)\Big]\,
\ln \Big(\frac{m^2 + V^{\prime\prime}}{\mu^2}\Big)\,,
\nonumber
\eeq
where the density of the cosmological constant term
$\rho_\La$ is included for completeness and \ $\mu^2$ is
the dimensional parameter of renormalization.

The evaluation of $Z(\phi)$ requires variable fields. The idea of
the standard procedure \cite{iliopoulos1975,fraser1985} is to split
the scalar field into a background constant $\phi_0$ and the variable
part $\ph$ as $\phi(x)= \phi_0 + \ph(x)$,  and evaluate the
effective action up to the second order in derivatives.
Since $\pa\phi(x)= \pa\ph(x)$, this approach gives a complete
result for $Z$.  Although one can apply this method to calculate
terms of the higher orders in derivatives \cite{fraser1985}, for
small oscillations near the minimum, these
terms are less relevant and we do not consider them here.

\subsection{Single scalar field in flat space}
\label{sec21}

To illustrate the procedure, we first perform the flat-space
calculation in the classical theory of a real scalar field with
the self-interaction $\la\phi^4$ . The Lagrangian density is
\beq
\mathcal{L}
\,=\,
\frac{1}{2}\,\pa_\mu \phi\, \pa^\mu \phi
\, - \, \frac{m^2}{2}\phi^2-\frac{\la}{4!}\,\phi^4.
\label{Lagrange}
\eeq
After Wick rotation,
the one-loop contribution to the effective action is given by
\beq
\Gamma^{(1)}(\phi)= \frac{1}{2}\text{Tr ln}(\Box+M^2),
\label{eq77}
\eeq
where
\beq
M^2= m^2+\frac{\lambda}{2}\phi^2(x).
\label{M2}
\eeq
In the next subsection, we consider the generalization to curved
spacetime, where
\beq
\Ga^{(1)}(\phi)
\,=\, \int d^4x \sqrt{-g}\,\,\Big\{
-\,V^{(1)}(\phi) \,+\, \frac{1}{2}\,Z^{(1)}(\phi)
\,g^{\mu\nu} \,\pa_\mu \phi\,
\pa_\nu \phi \,+\, ...\Big\}\,.
\label{Ga1}
\eeq

For now, consider the one-loop version of (\ref{EffPot}) in flat
space \cite{fraser1985}. To find the one-loop effective potential
\ $V^{(1)}$ \ it is
sufficient to consider a constant scalar $\phi_0$ in Eq.~(\ref{eq77}).
On the other hand, the derivation of $Z^{(1)}(\phi)$ requires setting
$\phi(x)=\phi_0 + \ph(x)$, with a subsequent expansion to the second
order in the coordinate-dependent part $\ph(x)$. The coefficients of
this expansion depend only on $\phi_0$. In this way, we obtain
\beq
&&
\Gamma^{(1)}(\phi_0+ \ph)
\,\,=\,\,
\int d^4x \,\,\Big\{ -\,V^{(1)}(\phi_0)\,
-\,\frac{\pa V^{(1)}}{\pa \phi_0}\,\ph
\,- \,\frac{1}{2}\,\frac{\pa^2 V^{(1)}}{\pa \phi_0^2}\,\ph^2
\nn
\\
&&
\qquad \qquad \qquad \qquad \qquad \qquad \qquad\quad
+\,\,\frac12\,Z^{(1)}(\phi_0)\,\pa_\mu \ph\,\pa^\mu \ph
\,+\,...\Big\}\,.
\label{eq74}
\eeq
To elaborate the last term, one has to expand the logarithm in
Eq.~(\ref{eq77})
\beq
&&
\Gamma^{(1)}(\phi_0+ \varphi)\,-\,\Gamma^{(1)}(\phi_0)
\,\,=\,\,
 \,\frac{\la}{4} \Tr \Big[
\frac{1}{p^2 + M_0^2}\big(2\phi_0 \varphi + \varphi^2\big)\Big]
\nn
\\
&&
\qquad \qquad \qquad \qquad
\,-\,
\frac{\lambda^2 \phi_0^2}{4} \Tr \Big[ \frac{1}{p^2 +  M_0^2}
\, \ph \, \frac{1}{p^2 +  M_0^2}\, \ph \Big] \,+\,...
\label{eq66pre}
\eeq
and commute all the momentum operators to the left using the
identity \cite{iliopoulos1975,fraser1985}
\beq
&&
\Phi \,\frac{1}{p^2 + M^2}
\,\,=\,\,
\frac{1}{p^2 + M^2}\,\Phi
\,+\, \frac{1}{(p^2 + M^2)^2}\,\big[p^2,\Phi\big]
\,+\, \frac{1}{(p^2 + M^2)^3}\,\big[p^2,\big[p^2, \Phi\big]\big]
\nn
\\
&&
\qquad
\qquad
\qquad
+\,\,\frac{1}{(p^2 + M^2)^4}\,
\big[p^2,\big[p^2,\big[p^2,\phi\big]\big]\big]\,\, + ...\,\,.
\label{eq27}
\eeq
The commutators of the momentum operator with an arbitrary function
of coordinates $\Phi$ follows the rules
(the four-dimensional momentum operator is identified
as $\,p_\mu = - i \pa_\mu$)
\beq
&&
\big[p^2, \Phi \big] \,=\,
\Box \Phi + 2ip^\mu \partial_\mu \Phi,
\nn
\\
&&
\big[p^2,\big[p^2, \Phi \big]\big]
\,=\,
\Box^2 \Phi \,+\,
4 i p^\mu \partial_\mu \Box \Phi
\,-\, 4 p^\mu p^\nu \partial_\mu \partial_\nu \Phi\,.
\label{eq34}
\eeq
After the described procedure we arrive at
\beq
&&
\Gamma^{(1)}(\phi_0+\varphi)
\,-\, \Gamma^{(1)}(\phi_0)
\,\, = \,\,
\frac{\lambda\phi_0}{2}\Tr \Big[ \frac{1}{p^2+M_0^2}\, \ph\Big]
 \, + \,
\frac{\lambda}{4}\, \Tr \Big[ \frac{1}{p^2 + M_0^2}\, \ph^2 \Big]
\nn
\\
&&
\qquad
\qquad
\qquad
- \,\,
\frac{\lambda^2 \phi_0^2}{4}\,
\Tr \Big[ \frac{1}{(p^2 + M_0^2)^2} \,\ph^2 \Big]
- \frac{\lambda^2 \phi_0^2}{4} \,\Tr
\Big[ M_0^2 \frac{1}{(p^2 + M_0^2)^4} \ph \Box \ph \Big].
\qquad
\label{commuta}
\eeq
Compared to Eq.~(\ref{eq74}), the first three terms in the \textit{r.h.s.}
are the derivatives of $V^{(1)}$. Concerning the kinetic part, we need
the last term with the d'Alembertian operator. The prescription for the
functional trace in momentum space can be found in \cite{fraser1985}.
In brief,
\beq
&&
\Tr \frac{1}{(p^2 + M^2)^n}
\,=\,
\int \frac{d^4 k}{(2\pi)^4} \frac{1}{(k^2 + M^2)^n}
\,=\, \frac{1}{(4\pi)^2 }
 \int_0^\infty \frac{2k^3 dk}{(k^2 + M^2)^n}.
 \label{proFraser}
 \eeq
Deriving the last integral for $n \geq 2$ is a simple exercise
 and we skip the details. After integrating by parts in the
last term of (\ref{commuta}), we arrive at
\beq
Z^{(1)}(\phi_0)
\,\,=\,\,
\frac{1}{12 (4 \pi)^2 }\,\frac{\la^2 \phi_0^ 2}{M^2}\,.
\label{eq78}
\eeq
As expected from the power counting, this expression is finite and
hence is independent of the renormalization procedure.

\subsection{Effective action of a single scalar field in curved space}
\label{sec22}

Consider now the generalization of the previous result to an
arbitrary curved spacetime. In this case, we have several options
for deriving the effective potential of the scalar field and
derivative expansion. The pioneer calculation of \cite{Shore1979}
used a technique specific for the potential of a scalar field
in de-Sitter spacetime. Technically more simple is the integration
of the renormalization group equation \cite{Buchbinder1985},
can be generalized for other sectors of effective action
\cite{Buchbinder1988} but since this approach is essentially tied
to the Minimal Subtraction scheme of renormalization, it is not
appropriate to explore the low-energy decoupling. The same
concerns the recently proposed approach of deriving effective
potential from the integration of trace anomaly \cite{Asorey2022}.
The most complete result for the massive scalar field can be
achieved by evaluating the form factors in the effective action
\cite{Peixoto2003}. However, the results for the heat-kernel
solution \cite{bavi90} are elaborated  explicitly  only for
the second-order terms \cite{apco,Omar-FF4D}. This technique
enables one to evaluate the $ (\pa \phi)^2$-type and $R\phi^2$-type
terms, but the  $R(\pa \phi)^2$-type corrections, especially in the
theory with two or more different masses, would require elaborating
on the heat-kernel solution in the third order in ``curvatures''
\cite{Barvinsky1990-34}. Thus, we need an appropriate method for
the effective potential and the derivative expansion in curved
spacetime. The mass-dependent scheme for the curved-space
effective potential has been developed using Riemann normal
coordinates and local momentum representation \cite{Bunch1979}.
The corresponding calculation can be found in \cite{CorPot} for a
general interaction term, so we only need to work out the
derivative expansion.

Let us derive the kinetic term $Z$ for the one-component scalar
field theory with a $\lambda \phi^4$ interaction term, following the
procedure which is a direct generalization of the one outlined in
the previous subsection \ref{sec21}. The action under consideration,
including the nonminimal term of the first order in curvature, has
the form
\beq
S(\phi, g_{\mu\nu})
= \int d^4x \, \sqrt{-g} \,
\Big\{
\frac12\, g^{\mu\nu} \pa_\mu \phi \pa_\nu \phi
- \frac12 \,\phi^2 m^2 +  \frac12 \,\xi R \phi^2
- \frac{\la}{4!} \phi^4 \Big\}.
\label{eq63}
\eeq
For the sake of simplicity, we choose the most common classical
potential $V(\phi) = \la \phi^4/4!$, different from the case of
effective potential in (\ref{REN-potya}).

The propagator expansion in the local momentum representation
can be done in different forms \cite{ Bunch1979}, and the one which
is the most appropriate for deriving the effective potential is, in the
Euclidean signature,\footnote{This expression
includes the normal coordinate expansion of $\sqrt{-g}$, such that
no corrections in the vertex part are required.}
\beq
\bar{G}(y)
\,=\, \int \frac{d^4k}{(2\pi)^4} \, e^{iky}
\bigg[
\frac{1}{p^2 + M_0^2} \,-\,
\frac{\tilde{\xi} R}{(p^2 + M_0^2)^2} \bigg],
\label{eq 64}
\eeq
where
\beq
M_0^2 \,=\, m^2 + \frac{\la}{2}\,\phi^2_0
\qquad
\mbox{and}
\qquad
\tilde{\xi} = \xi - \frac16.
\eeq
Let us note that since we restrict our attention to the
\ $\mathcal{O}(R)$ \ terms, hence $R$ can be treated as a constant.
The method can be applied beyond this approximation, but this
would require more calculations.

Following the same steps as in the flat case, but
with the propagator (\ref{eq 64}), we obtain
\beq
&&
\Ga(\phi_0 + \varphi, \,g_{\mu\nu}) - \Ga(\phi_0, g_{\mu\nu})
\nn
\\
&&
\qquad
\,\,=\,\, \frac{1}{2} \Tr \, \ln \bigg\{
1 + \frac{\la}{2}
\Big( \frac{1}{k^2 + M_0^2}
- \,\,\frac{\tilde{\xi} R}{(k^2 + M_0^2)^2}\Big)
\big(2\phi_0 \varphi + \varphi^2\big) \bigg\}.
\label{eq65}
\eeq
Expanding the logarithm into series, discarding terms of the second
order in curvatures, and also higher orders in $\phi$, we arrive at
the expression
\beq
&&
\Ga(\phi_0 + \varphi, \,g_{\mu\nu}) - \Ga(\phi_0, g_{\mu\nu})
\nn
\\
&&
\qquad
\,\,=\,\,
\frac{\la}{4} \Tr \Big[
\frac{1}{p^2 + M_0^2}\big(2\phi_0 \varphi + \varphi^2\big)\Big]
- \,\frac{\la}{4}\,\tilde{\xi} R\, \Tr
\Big[\frac{1}{(p^2 +  M_0^2)^2} (2\phi_0 \varphi + \varphi^2)\Big]
\label{eq66}
\\
 &&
\qquad
-\,
\frac{\lambda^2 \phi_0^2}{4} \Tr \Big[ \frac{1}{p^2 +  M_0^2}
\, \ph \, \frac{1}{p^2 +  M_0^2}\, \ph \Big]
\, + \,
 \frac{\la^2}{2}  \,\tilde{\xi} R  \phi_0^2\,
 \Tr \Big[
 \frac{1}{p^2+ M_0^2}\,\ph\,\frac{1}{(p^2 +  M_0^2)^2}\,\ph\Big].
\quad
\nn
\eeq
As we already know, the kinetic term arises because of the commutation
of $\ph$ with the momentum operator. Therefore, we can focus only
on the last two terms of Eq.~(\ref{eq66}).

It proves useful to express the coefficient $Z^{(1)}(\phi)$ as
\beq
Z^{(1)}(\phi) = Z^{(1)}_0 + Z^{(1)}_1 R + \mathcal{O}(R^2)\,.
\label{eq68}
\eeq
The first term, $Z^{(1)}_0$, corresponds to the flat space and has
already been calculated in (\ref{eq78}). One can rewrite it by
replacing $M_0^2 \longrightarrow M^2$,
\beq
Z^{(1)}_0 = \frac{1}{12} \frac{\lambda^2 \phi_0^2}{16 \pi^2 M^2}
\label{eq69}
\eeq

The last term in (\ref{eq66}), after commutations, becomes
\beq
&&
\frac{ \la^2}{2}\,\tilde{\xi} \phi_0^2\, R \,\Tr
\Big[
\frac{1}{p^2 +M_0^2} \,\ph\, \frac{1}{(p^2 + M_0^2)^2}\, \ph
\Big]
\,\,= \,\,
\frac{ \la^2}{2}\, \tilde{\xi} \phi_0^2\,  R \,
\Tr \Big\{
\frac{1}{(p^2 + M_0^2)^3}\, \ph^2
\nn
\\
&&
\qquad
\qquad
\qquad
+\,\, \frac{1}{(p^2 + M_0^2)^4} \big[p^2, \ph\big] \ph
\,+\,
\frac{1}{(p^2 + M_0^2)^5} \big[p^2, \big[p^2, \ph\big]\big] \ph
+ \,... \Big\}.
\label{eq70}
\eeq
The first term of the last expression contributes to the second
derivative of the effective potential because
\beq
&&
\frac{\la}{4} \Tr \Big[
\frac{1}{p^2 + M_0^2} \varphi^2\Big]- \,\frac{\la}{4}\,\tilde{\xi} R\, \Tr
\Big[
\frac{1}{(p^2 +  M_0^2)^2}  \varphi^2
\Big]
\,-\,
\frac{\lambda^2 \phi_0^2}{4} \Tr \Big[ \frac{1}{(p^2 +  M_0^2)^2}
\, \ph^2 \,\Big]
\nn
\\
&&
\qquad
\qquad
\qquad
\,+\,\,
\frac{\la^2 \tilde{\xi}\phi_0^2}{4} \Tr \Big[ \frac{1}{(p^2 +  M_0^2)^3}
\, \ph^2 \,\Big]
\,=\, \frac12 \int d^4x \, \frac{\pa^2 V^{(1)}}{\pa \phi_0^2} \varphi^2\,.
\label{formerblue}
\eeq
After integrating by parts, we obtain the first-order correction
to (\ref{eq78}) in the form
\beq
Z^{(1)}_1
\,\,=\, \, \frac{\la^2}{24 (4\pi)^2 M_0^4}
\,\,\tilde{\xi} R \phi_0^2\,.
\label{eq72}
\eeq
Replacing $\phi_0^2$ by $\phi^2$, the coefficient  $Z^{(1)}(\phi)$,
in the $\mathcal{O}(R)$ approximation, is given by
\beq
Z^{(1)}\,\,=\,\, \frac{1}{12\,(4\pi)^2}\,
\, \frac{\la^2 \phi^2}{M^2}\,
\bigg(1 \,+\, \frac{\tilde{\xi}R}{2 M^2} \bigg).
\label{eq73}
\eeq
Remarkably, this expression does not depend on the renormalization
parameter $\mu$. This is explained by the fact that, in the pure
scalar theory, at the one-loop level, there are no divergences in
the kinetic term.

At this point it is worthwhile to make a
clarification about the approximations used in the formula
(\ref{eq73}). In flat spacetime, the derivative expansion
(\ref{EffPot}) assumes that the derivatives of the background
scalar are subleading compared to the potential part, i.e.,
$\big(\pa \phi\big) \ll M^4$ or, in the detailed form,
$\big(\pa \phi\big) \ll \la \phi^4 + \phi^2m^2$. It is important to
note that $M^2$ depends on the background scalar, such that
the definition (\ref{EffPot}) and all consequent formulas
hold even in the massless limit \cite{ColeWein,Jackiw74}.
The mentioned approximation remains necessary in the next
orders of expansion \cite{fraser1985},
and also in the models with fermions and/or vectors, where
the kinetic term is divergent and gains logarithmic corrections.
This is the limitation behind Eq.~(\ref{eq78}), but the curvature
correction in (\ref{eq73}) implies also the expansion in the
Riemann normal coordinates, which means an approximation
of weak gravitational field. In our setting, this means
$\big| R_{\mu\nu\al\be}\big| \ll M^2$, that have to hold simultaneous
with the aforementioned flat-space restriction
$\big(\pa \phi\big) \ll M^4$.

\section{The case of 
a two-component scalar field}
\label{sec3}

Consider a model with two real interacting scalar fields $\phi$
and $\chi$, described by the covariant Lagrangian density
\beq
&&
\mathcal{L}
\,=\, \frac12\,g^{\mu\nu} \pa_\mu \phi \pa_\nu \phi
+ \frac12\,g^{\mu\nu} \pa_\mu \chi \pa_\nu \chi
- \frac12\,m^2\phi^2 - \frac12\,M^2\chi^2
\nn
\\
&&
\qquad\quad
+ \,\,\frac12\,\xi_1 R\phi^2 + \frac12\,\xi_2 R\chi^2
- \frac{\lambda_1}{4!}\phi^4
- \frac{\lambda_2\chi^4}{4!}
- \frac{\lambda_{12}}{8}\phi^2\chi^2\,.
\label{eq1}
\eeq
Here $m$, $M$ are the masses of the two fields
and $\la_1$, $\la_2$, $\la_{12}$ are the coupling constants.

In what follows, we show how to reduce the quantum calculations in
the theory with two masses to the single-mass case. Our approach is
based on the background field method and the change of variables
(diagonalization) which was previously used in \cite{Buchbinder2019}
(see also \cite{Tensors} for a general discussion of diagonalization
and \cite{Petrov} for the references to original papers and more
profound discussion, including the non-positively defined oscillators).
The rest of the procedure is analogous to using normal modes in the
multidimensional oscillator.

\subsection{Diagonalization of the bilinear form of the action}

Following the background field method, we decompose scalar fields
into background $\phi$, $\chi$ and quantum  $\si_1$, $\si_2$
counterparts as
\beq
\phi \,\longrightarrow \,\phi+\si_1\,,
\qquad
\chi \,\longrightarrow\,\chi+\si_2\,.
\label{bfm}
\eeq
The one-loop contribution depends on the bilinear in quantum fields
action
\beq
S^{(2)} \,\,=\,\, \frac{1}{2} \int d^4x \, \sqrt{-g} \,\,
\big(
\begin{array}{cc}
\si_1 & \si_2
\end{array}
\big)
\hat{H}
\left(
\begin{array}{c}
\si_1 \\
\si_2
\end{array}
\right)\,.
\label{eq2}
\eeq
Operator $\hat{H}$ has a standard matrix form
\beq
\hat{H} \,=\,
\left(
\begin{array}{cc}
H_{11} & H_{12} \\
H_{21} & H_{22}
\end{array}
\right)
\,=\,
\hat{1} \Box +
\left( \begin{array}{cc}
    M_{11}^2 & M_{12}^2\\
    M_{21}^2 & M_{22}^2
\end{array}
\right)\,,
\label{eq3}
\eeq
where
\beq
\qquad
\hat{1} \,=\,
\left(
\begin{array}{cc}
   1 & 0 \\
   0 & 1 \end{array}
\right)
\qquad
\mbox{and}
\qquad
\left\{
\begin{array}{c}
M_{11}^2 = \, m^2 - \xi_1R+  \frac{\la_1}{2}\,\phi^2
+ \frac{\la_{12}}{4} \chi^2,
\\
M_{12}^2 =  M_{21}^2 = \,\la_{12} \,\phi \chi,
\qquad
\,\,\quad
\qquad
\\
M_{22}^2
=  \, M^2 - \xi_2R + \frac{\la_2}{2} \chi^2 + \frac{\la_{12}}{4} \phi^2\,.
\end{array}
\right.
\label{Ms}
\eeq
To diagonalize matrix $\hat{H}$, we perform rotation to the
angle $\al$ in the space of the fields,
\beq
\left(
\begin{array}{c}
\si_1 \\
\si_2
\end{array}
\right)
\,=\,
\left(
\begin{array}{cc}
\cos \alpha & -\sin \alpha \\
\sin \alpha & \,\,\cos \alpha
\end{array}
\right)
\left(
\begin{array}{c}
\rho_1 \\
\rho_2
\end{array}
\right) .
\label{eq4}
\eeq
This transformation preserves the diagonal form of the derivative
term in (\ref{eq3}) and, on the other hand, its Jacobian equals to
one. After rotation, the bilinear action becomes
\beq
&&
S^{(2)}
\,=\,
\frac{1}{2} \int d^4x \sqrt{-g}  \Big\{ \rho_1\Box\rho_1
+\rho_2\Box\rho_2
\nn
\\
&&
\qquad \qquad
+ \,\,
\rho_1 \big(  M_{11}^2\cos^2 \alpha
+ 2 M_{12}^2 \sin \alpha \cos \alpha
+  M_{22}^2\sin^2 \alpha\big) \rho_1
\nn
\\
&&
\qquad \qquad
+ \,\,
 \rho_2 \big( M_{11}^2 \sin^2 \al
- 2 M_{12}^2 \sin \alpha \cos \al
+ M_{22}^2 \cos^2 \alpha  \big) \rho_2
\nn
\\
&&
\qquad \qquad
+ \,\,
\rho_1  \big[ (M_{22}^2
- M_{11}^2) \sin 2 \al
\, + \,
2 M_{12}^2 \cos 2 \al\big] \rho_2 \Big\}\,.
\label{eq5}
\eeq
Choosing
\beq
 \theta \,=\, \cot 2\al \,=\,
\frac{M^2_{22} - M^2_{11} }{2M_{12}^2}
\label{eq7}
\eeq
we eliminate the non-diagonal components. It is useful to introduce
new notations
\beq
&&
a =\cos^2 \alpha  =  \frac{1}{2}+\frac{\theta}{2\sqrt{1+\theta^2}}
\,,\qquad
b = \sin^2\alpha  =  \frac{1}{2}-\frac{\theta}{2\sqrt{1+\theta^2}}\,,
\nn
\\
&&
c  =\sin 2\al =  \frac{1}{\sqrt{ 1+ \theta^2}}.
\label{eq10}
\eeq

In the new variables, the bilinear form of the action becomes
\beq
\hat{\mathcal{H}}
\,=\,
\left(
\begin{array}{cc}
\Box +\Pi_1 & 0 \\
0 & \Box + \Pi_2
\end{array}
\right)
\,=\,
\left(
\begin{array}{cc}
\mathcal{H}_1 & 0 \\
0 & \mathcal{H}_2
\end{array}
\right),
\label{eq12}
\eeq
where
\beq
&&
\Pi_1 =  a M_{11}^2 + bM_{22}^2 -c M^2_{12}\,,
\nn
\\
&&
\Pi_2 =  b M_{11}^2 +aM_{22}^2 + c M^2_{12}\,.
\label{PiPi}
\eeq
At this point, the diagonalization is concluded. The main difference
with the case of two independent scalar fields is that each of the
two expressions $\Pi_1$ and $\Pi_2$ depend on both background
scalars $\phi$ and $\chi$. In what follows, we show how this
difference can be eliminated.

\subsection{Setting the framework}
\label{sec3.1}

The one-loop effective action is proportional to the expression that
can be factorized according to Eq.~(\ref{eq12}), using the unity of
the Jacobian of (\ref{eq4}),
\beq
\ln\Det \hat{H}
\,=\, \ln\Det \hat{\mathcal{H}}
\,=\,  \Tr \ln \hat{\mathcal{H}}
\,=\, \Tr\ln \mathcal{H}_1
\,+\,
\Tr\ln \mathcal{H}_2\,.
\label{eq17}
\eeq
Thus, the one-loop contribution to the effective action is
\beq
\Ga^{(1)}(\phi,\chi)
\,=\,
\frac{1}{2} \Tr \ln (\Box +\Pi_1)
+
\frac{1}{2} \Tr \ln (\Box +\Pi_2)\,.
\label{eq18}
\eeq

To employ the method described in the previous section, let us
first perform the splitting of the scalar fields into constant and
weakly variable parts,
\beq
\phi \,\longrightarrow \,\phi_0 + \ph(x)
\quad \,\,
\mbox{and}
\,\, \quad
\chi \,\longrightarrow \,\chi_0 + \vp(x).
\label{bfm2}
\eeq

The desired derivative expansion up to the second order in
the scalar fields $\ph$ and $\chi$ is
\beq
&&
\Ga^{(1)}(\phi,\chi)
\,=\,
\int d^4x \sqrt{-g} \Big\{
- V^{(1)}(\phi,\chi)
+ \frac12\,Z_\phi^{(1)} (\pa \ph)^2
+ \frac12\, Z_\chi^{(1)} (\pa \vp)^2
\nn
\\
&&
\qquad
\qquad
\qquad
+ \,\,Z_{\phi\chi}^{(1)} (\pa \ph)(\pa \ep)
\,+\, ...
\Big\},
\label{eq76}
\eeq
where we used compact notations
\beq
(\pa \ph)^2 \,=\, g^{\mu\nu} \pa_\mu \phi \,\pa_\nu \phi,
\qquad
(\pa \vp)^2 \,=\, g^{\mu\nu} \pa_\mu \vp \,\pa_\nu \vp,
\qquad
(\pa \ph)(\pa \vp) \,=\, g^{\mu\nu} \pa_\mu \phi \,\pa_\nu \vp\,.
\eeq
In the rest of this section, we first present direct calculation of
the coefficients $Z^{(1)}$ and then report about the method of
reduction to the single scalar case.

\subsection{Direct calculation of the derivative expansion}
\label{3.3}

Without the loss of generality, one can assume that the coefficients
$Z_\phi^{(1)}$, $Z_\chi^{(1)}$ and $Z_{\phi\chi}^{(1)}$ do not
depend on the variable scalar fields, but only on the constant
components of the scalars, couplings, masses and nonminimal
parameters \ $\xi_{1,2}$.

Since we need only the terms quadratic in the variable components,
let us expand the effective action into the power series up to the
second order in $\ph$ and $\vp$,
\beq
&&
\Ga^{(1)}(\phi_0 + \ph, \chi_0 + \vp)
\,-\, \Gamma^{(1)}(\phi_0, \chi_0)
\,=\,
\int d^4x \bigg\{  -\frac{\pa V^{(1)}(\phi,\chi)}{\pa \phi_0} \ph
- \,\frac{\pa V^{(1)}(\phi,\chi)}{\pa \chi_0} \vp
\nn
\\
&&
\qquad \qquad \qquad
-\, \frac12\,\frac{\pa^2 V^{(1)}(\phi,\chi)}{\pa \phi_0^2}\, \ph^2
- \frac12 \, \frac{\pa^2 V^{(1)}(\phi,\chi)}{\pa \chi_0^2} \,\vp^2
- \frac{\pa^2 V^{(1)}(\phi, \chi)}{\pa \phi \pa\chi} \,\ph \vp
\nn
\\
&&
\qquad\qquad \qquad\qquad
+\,  \frac12\, Z_\phi^{(1)} (\pa \ph)^2
+ \frac12\, Z_\chi^{(1)} (\pa \vp)^2
        + Z_{\phi\chi}^{(1)} (\pa \ph)(\pa \vp) + \dots \bigg\}.
\label{eq20}
\eeq

On the other hand, we have to expand $\Pi_{j}$
in Eq.~(\ref{eq18}) as follows:
\beq
&&
\Pi_{j}(\phi_0+\ph; \chi_0 +\vp)
\,=\, \Pi_{j0}(\phi_0,\chi_0)
+ \frac{\pa\Pi_{j}}{\pa\phi_0}\,\ph
+ \frac{\pa\Pi_{j}}{\pa\chi_0}\,\vp
\nn
\\
&&
\qquad \qquad \qquad \qquad \quad
+ \,\,\frac12\,\frac{\pa^2\Pi_{j}}{\pa\phi_0^2}\,\ph^2
+ \frac12\, \frac{\pa^2\Pi_{j}}{\pa\chi_0^2}\,\vp^2
+ \frac{\pa^2\Pi_{j}}{\pa\phi_0 \pa\chi_0}\,\ph\vp +...
\label{eq80}
\eeq
As in the one-scalar case, the terms involving second derivatives do
not contribute to the kinetic term. Thus, we need to keep only the
terms of the first order in derivatives. Furthermore, in the two
first-derivative terms, the coefficients $\pa\Pi_{j}/\pa\phi_0$ and
$\pa\Pi_{j}/\pa\chi_0$ depend solely on the constants $\phi_0$ and
$\chi_0$. That being said, we can consider that
\beq
\frac{\pa\Pi_{j}}{\pa\chi_0} \,=\, \Pi_{\chi}(\chi_0)\,=\, \Pi_{\chi}
\qquad \mbox{and} \qquad
\frac{\pa\Pi_{j}}{\pa\phi_0}\,=\, \Pi_{\phi}(\phi_0)\,=\, \Pi_{\phi}
\label{Piphchi}
\eeq
are constant coefficients of the linear expansion
\beq
&&
\Pi_{j}(\phi_0+\ph; \chi_0 +\vp)
\,=\, \Pi_{0j}(\phi_0,\chi_0)
+  \Pi_{\phi j} \ph
+  \Pi_{\chi j}\vp\,.
\label{eq80simp}
\eeq

For the effective potential (\ref{REN-potya}), it is sufficient to
set $\phi=\phi_0$ and $\chi=\chi_0$ in (\ref{eq18}),
\beq
V^{(1)}
\,=\, -\frac{1}{2} \int \frac{d^4k}{(2\pi)^4}
\Big[\ln \big(k^2+\Pi_{01}\big) \,+\, \ln\big(k^2 +\Pi_{02}\big)\Big]\,.
\label{eq23}
\eeq
The lowest-order derivative terms follow from the relevant
part of (\ref{eq80}), so we get
\beq
\Ga^{(1)} \,+\, \int d^4x\, V^{(1)}
 \,\,=\,\,
\frac{1}{2} \sum_{j=1,2}\,
\Tr \ln \Big[1\,+\,
\frac{1}{p^2 + \Pi_{0j}} \big(\Pi_{\phi j}\ph+\Pi_{\chi j}\vp\big)
\Big]\,.
\label{eq24}
\eeq
From this point onward  the arguments will be suppressed,
assuming
\beq
\Ga^{(1)}(\phi_0 + \ph, \chi_0 + \vp) =\Ga^{(1)}
\qquad \mbox{and} \qquad
\Pi_{0j}= \Pi_0(\phi=\phi_0, \chi=\chi_0)\,,
\label{eq21}
\eeq
such that, for each \ $j=1,2$, \ in the first order of the power
series, we get (here $\Pi_{\phi j } \to \Pi_{\phi }$, \ $\Pi_{\chi j}
\to \Pi_{\chi }$ and $ \Pi_{0j}\to \Pi_{0}$ for brevity)
\beq
&&
\bar{\Ga}_j \,\,=\,\,
\frac{1}{2} \Tr \ln
\bigg[1 + \frac{1}{p^2 + \Pi_{0}}\,
\big(\Pi_{\phi }\ph + \Pi_{\chi }\vp\big)
\bigg]
\,=\, \,\frac{1}{2}\Tr \bigg[
\frac{1}{p^2 + \Pi_{0}}\,\big(\Pi_{\phi }\ph + \Pi_{\chi }\vp\big)
\nn
\\
&&
\qquad
-\,\, \frac12\, \frac{1}{p^2 + \Pi_{0}} \big(\Pi_{\phi}\ph
+ \Pi_{\chi}\vp\big) \,\frac{1}{p^2 + \Pi_{0}}\,
\big(\Pi_{\phi}\ph+\Pi_{\chi}\vp)\, +... \bigg]\,.
\label{eq26}
\eeq
In the second term of the \textit{r.h.s.} of the last relation,
we commute all of the momentum operators to the left of the
field functions, in the way analogous to Eq.~(\ref{eq27}). Then
\beq
&&
\bar{\Ga}_j \,\,=\,\,
 \frac{1}{2} \Tr \bigg\{
\frac{ 1}{p^2 + \Pi_0}\,\big(\Pi_{\phi }\ph+\Pi_{\chi }\vp\big)
- \,\frac12\,  \frac{1}{( p^2 + \Pi_0)^2}  (\Pi_{\phi }\ph
+ \Pi_{\chi }\vp)^2
 \nn
 \\
 &&
 \qquad  \qquad
- \,\, \frac{\Pi_\phi^2}{2}\frac{1}{(p^2 + \Pi_0)^3}\big[p^2,\ph\big]\ph
-  \frac{\Pi_\phi^2}{2}   \frac{1}{(p^2 + \Pi_0)^3}  \big[p^2,\ph\big]\vp
 \nn
 \\
 &&
 \qquad  \qquad
 - \,\,
\Pi_\chi\Pi_\phi   \frac{1}{(p^2 + \Pi_0)^3}
\big[p^2,\vp\big]\ph
-  \frac{\Pi_\phi^2}{2}   \frac{1}{(p^2 + \Pi_0)^4}
 \big[p^2,\big[p^2,\ph\big]\big]\ph
 \nn
 \\
 &&
 \qquad  \qquad
 - \,\,
\frac{\Pi_\phi^2}{2}   \frac{1}{(p^2 + \Pi_0)^4}
\big[p^2,\big[p^2,\ph\big]\big]\vp
- \Pi_\chi\Pi_\phi  \frac{1}{(p^2 + \Pi_0)^4}
\big[p^2,\big[p^2,\vp\big]\big]\ph
 +\,... \bigg\}.
 \qquad
\label{eq28}
\eeq
The part of effective action which is linear in $\phi$ and $\chi$,
corresponds to the first term of last expression, i.e.,
\beq
&&
\int d^4x \,\, \bigg(\frac{\pa V^{(1)}}{\pa \phi_0} \ph
+ \frac{\pa V^{(1)}}{\pa \chi_0}\vp\bigg)
\,\,=\,\,
\frac{1}{2} \sum_{j=1,2}
\Tr  \,\frac{1}{p^2 + \Pi_{0j}} \,
\big(\Pi_{\phi j}\ph + \Pi_{\chi j}\vp\big)
\nn
\\
&&
\qquad
\qquad
\qquad
\qquad
=\,\,\frac12\,\sum_{j=1,2}
\int d^4x  \int \frac{d^4k}{(2\pi)^4}
\,\frac{1}{k^2+\Pi_{0j}}\,\big(\Pi_{\phi j}\ph+\Pi_{\chi j }\vp\big)\,.
\label{eq29}
\eeq

Similarly, the second term of Eq.~(\ref{eq28}) produces
the quadratic terms
\beq
&&
\int d^4x \,\bigg(\frac12\,\frac{\pa^2 V^{(1)}}{\pa \phi_0} \ph^2
+ \frac12\,\frac{\pa^2 V^{(1)}}{\pa \chi_0} \vp^2
+\frac{\pa^2V^{(1)}}{\pa\phi_0\pa\chi_0}\,\ph\vp\bigg)
\nn
\\
&&
\qquad
=\,\,
\frac{1}{4}\,\sum_{j=1,2}
\int d^4x \int \frac{d^4k}{(2\pi)^4}\,
\frac{1}{( p^2 + \Pi_{0j})^2}\,
\big(\Pi_{\phi j}\ph+\Pi_{\chi j}\vp\big)^2\,.
\label{eq31}
\eeq

Now we are in the position to derive the coefficients $Z^{(1)}_{\ph}$,
$Z^{(1)}_{\chi}$ and $Z^{(1)}_{\ph\chi}$ of derivative expansion
from Eq.~(\ref{eq76}).
The respective terms with two derivatives appear as a result of the
commutation of the momentum operators with the functions of the
coordinates in Eq.~(\ref{eq28}). In the framework of local momentum
representation, the commutators are the same (\ref{eq34}) as in the
flat spacetime. Commuting, taking the traces, and integrating over
the momenta, we find, for each component \ $\Pi_j$,
\beq
-\, \frac{1}{24 (4\pi)^2}\, \frac{1}{\Pi_0}
\int d^4x \,\,
\big(\Pi_\phi^2 \,\ph\Box\ph \,+\, \Pi^2_\chi \,\vp\Box\vp
\,+\,2\Pi_\phi \Pi_\chi \,\vp \Box \ph\big).
\label{eq38}
\eeq
As explained above, when integrating by parts, one can regards
the coefficients $\Pi_\phi$ and $\Pi_\chi$ constants and get
\beq
&&
Z^{(1)}_1(\phi,\chi)\, = \,  \frac{1}{12 (4 \pi)^2 }
\bigg(\frac{\Pi_{\phi1}^2}{\Pi_{01}} +\frac{\Pi_{\phi2}^2}{\Pi_{02}}
\bigg),
\nn
\\
&&
Z^{(1)}_2(\phi,\chi)
\, = \,  \frac{1}{12 (4 \pi)^2 } \left(\frac{\Pi_{\chi1}^2}{\Pi_{01}} +\frac{\Pi_{\chi2}^2}{\Pi_{02}}\right),
\nn
\\
&&
Z^{(1)}_{12}(\phi,\chi) \, = \,
\frac{1}{12 (4 \pi)^2 } \left(\frac{\Pi_{\phi1}\Pi_{\chi_1}}{\Pi_{01}}
+\frac{\Pi_{\phi2}\Pi_{\chi_2}}{\Pi_{02}}\right)\,.
\label{eq45}
\eeq
Since the variable parts in the expressions for
$Z^{(1)}_{\ph}$, $Z^{(1)}_{\chi}$ and $Z^{(1)}_{\ph\chi}$
are irrelevant in the given approximation, one can simply
replace $\phi_0$ by $\phi(x)$ and $\chi_0$ by $\chi(x)$ in the
last formulas to arrive at the final answer. The explicit
expressions for the coefficients (\ref{eq45}) are rather bulky
and cannot be presented here. The coefficients are finite as
predicted by the power counting and hence independent of
the renormalization condition or parameter $\mu$.

It is worth noting that the method of derivative expansion that
we used here can be applied for calculating higher-order derivative
terms, as it was done for the single scalar field in flat spacetime in
the work \cite{fraser1985}. On the other hand, the part which we
derived here is certainly dominating in all possible applications.

\subsection{Double diagonalization and reduction to the
single scalar model}
\label{3.4}

An alternative approach to derive the coefficients $Z^{(1)}$ involves
introducing an auxiliary field $\rho$ which is a linear combination
of $\ph$ and $\vp$, and expanding $\Pi_{j}$ in a single field
$\rho$, 
\beq
\Pi_{j}(\phi_0+\ph,\, \chi_0 +\vp)
\, \, = \, \,
\Pi_{0j}+\Pi_{\phi j}\ph+\Pi_{\chi j} \vp
\,\,  = \,\, \Pi_{0j} +\Pi_{\rho j} \rho ,
\label{double}
\eeq
where (for both $j=1,2$)
\beq
\Pi_{\rho}\, = \,\sqrt{\Pi_\phi^2+\Pi_\chi^2}
\qquad
\mbox{and}
\qquad
\rho \, = \, \frac{\Pi_\phi}{\Pi_\rho}\,\ph
\,+\,
\frac{\Pi_\chi}{\Pi_\rho}\,\vp\,.
\eeq
Introducing $\rho$ is analogous to what is called derivative along
the given direction in the analysis of many variables.
Together with the rotation (\ref{eq4}) the procedure can be called
double diagonalization because it includes rotation (\ref{eq4})
with the angle (\ref{eq7}) in the space of quantum fields and
the change of ``coordinates'' (\ref{double}) in the space of
background fields. Together, these two operations reduce the
calculation in the two (and more, if necessary) scalar fields theory
to the single-scalar calculation. Let us now see how it works.

In the expansion
\ $\Pi(\phi_0+\ph,\, \chi_0 +\vp) \, = \, \Pi_{0} +\Pi_{\rho} \rho$ \
in Eq.~(\ref{double}), the coefficients $\Pi_0$ and $\Pi_{\rho}$
are constants. Thus, for each $j$, in the relevant order of the
expansion in derivatives,
\beq
&&   \Tr \ln \bigg[
1 + \frac{1}{p^2+\Pi_0}\,\big(\Pi_0+\Pi_{\rho}\rho\big)\bigg]
\nn
\\
&&
\qquad
  =\,\,
  \frac{1}{2}\, \Tr
\left[\Pi_{\rho}\,\frac{1}{p^2+\Pi_0}\, \rho (x)
\,-\, \frac{\Pi_{\rho}^2}{2}\,\frac{1}{p^2+\Pi_0}\,
\rho(x) \,\frac{1}{p^2 + \Pi_0} \rho(x)\right]\,.
\label{eq90}
\eeq
Taking traces and integrating by parts are essentially the same as
in Sec.~\ref{sec2}, so we do not need to repeat the details. After
that, one has to make a replacement according to (\ref{double}),
\beq
&&
\frac{1}{24(4\pi)^2}\,\frac{\Pi_{\rho}^2}{\Pi_0}
\int d^4x \,(\pa_\mu \rho)(\pa^\mu \rho)
\,\,= \,\,
\frac{1}{24(4\pi)^2 \,\Pi_0}
\int d^4x \Big[
\Pi_\phi^2 (\pa_\mu \ph)(\pa^\mu \ph)
\nonumber
\\
&&
\qquad
\qquad
\qquad
\,+\,\, \Pi_\chi^2 \,(\pa_\mu \vp)(\pa^\mu \vp)
\,+\, 2\Pi_\phi \Pi_\chi \,(\pa_\mu \ph)(\pa^\mu \vp) \Big]\,.
\label{eq91}
\eeq
Now, summing up the contribution of $\Pi_1$ e $\Pi_2$, we
arrive at the same result (\ref{eq45}). So, we can conclude that
the described double diagonalization is operational.

Using the described approach, we can easily obtain $Z^{(1)}$ in the
first order in curvature, simply reformulating the result (\ref{eq73})
from Sec.~\ref{sec2}, i.e.,
\beq
\frac{\tilde{\xi} R}{24 (4\pi)^2}
\, \frac{\Pi_{\rho}^2}{\Pi_0^2}
\, \,
\int d^4x \,\,(\pa_\mu \rho)(\pa^{\mu} \rho)\,.
\label{eq97}
\eeq
After the same replacement as in (\ref{eq91}), we get
\beq
&&
Z_{\phi}
\,=\, \frac{1}{12(4 \pi)^2} \bigg[ \Big(1
+ \frac{\tilde{\xi}_1 R}{2\Pi_{01}} \Big)
   \,\frac{\Pi_{\phi 1}^2}{\Pi_{01}}
+ \Big(1 + \frac{\tilde{\xi}_2 R}{2\Pi_{02}}\Big)
   \,\frac{\Pi_{\phi2}^2}{\Pi_{02}}
\bigg]\,,
\nn
\\
&&
Z_{\chi}\, = \,\frac{1}{12(4 \pi)^2}
\bigg[ \Big( 1 + \frac{\tilde{\xi}_1 R}{2\Pi_{01}} \Big)
\frac{\Pi_{\chi_1}^2}{\Pi_{01}}
+ \Big(1 + \frac{\tilde{\xi}_2 R}{2\Pi_{02}} \Big)
\frac{\Pi_{\chi_2}^2}{\Pi_{02}} \bigg]\,,
\nn
\\
&&
Z_{\chi \phi}\, = \,\frac{1}{12(4 \pi)^2}
\bigg[ \Big( 1 + \frac{\tilde{\xi}_1 R}{2\Pi_{01}} \Big)
\frac{\Pi_{\phi_1}\Pi_{\chi_1}}{\Pi_{01}}
+ \Big( 1 + \frac{\tilde{\xi}_2 R}{2\Pi_{02}} \Big)
\frac{\Pi_{\phi_2}\Pi_{\chi_2}}{\Pi_{02}}\bigg]\,.
\label{eq101}
\eeq

The described reduction of the multi-scalar calculation to the
single scalar case can be directly applied for a greater number of
real scalars. The number of directional
derivatives (\ref{double}) is always equal to the number of
original scalars and we do not expect a degeneracy in this
expansion, at least for most of the possible models with real
scalars. The application to the system of complex and real scalars
(e.g., Higgs field plus an inflaton) is also possible, but would
require writing a complex field as two real scalars and also, in
the case of the Standard Model Higgs, solving the problem of
group indices.

\section{Low-energy limit in the theory with
hierarchy of the masses }
\label{sec4}

The notion of quantum decoupling is an important element in the
effective approach to quantum field theory and quantum gravity.
According to the Appelquist and Carazzone theorem in QED \cite{AC},
quantum contribution of the field with a large mass $M$ vanish
as $k^2/M^2$ in the IR (low energy limit), where the Euclidean
momentum is small, i.e., $k^2 \ll M^2$. We can observe this effect
in many other situations, including for the polarization operator
of gravitons in the semiclassical gravity \cite{apco} (see more
references, e.g., in \cite{OUP}). The derivative expansion makes
sense in the situation when the scalar field performs small
oscillations near the point of the minima of the
potential. This situation fits pretty well with the decoupling
theorem setting, so one can ask how the decoupling can be observed
in the effective potential\footnote{It was known from the epoch
of Heisenberg and Euler \cite{EH-1938} that in the interaction
sector quantum corrections are qualitatively different in the UV
and in the IR.}
and in the first coefficients of the derivative
expansion. Indeed the oscillations may occur not in a simple
central minimum of the potential, but also in the broken phase of
the ``Mexican hat'' potential of the Spontaneous Symmetry Breaking.
In curved spacetime, this consideration is possible but meets
certain complications \cite{sponta,Tmn-sponta}. In the considerations
presented below, we consider only local motions in the vicinity of
the minimum of the potential, such that the origin of this minimum
does not matter and we may consider only the basic situation. For
this reason, we assume that the magnitude of the scalar field is
small when this field oscillates in the vicinity of the point of the
minimum.

The decoupling in the case of a single-field effective potential is
quite obvious and intuitive. The logarithms in the expression
(\ref{REN-potya}) for the theory with mass $M$, comes together
with the divergences, in the
same way as $\ln \big[ (k^2 + M^2)/\mu^2 \big]$-type formula
emerges\footnote{The real expression is more complicated
\cite{Peixoto2003}, but the difference with this simplified formula
is sub-logarithmic in the UV, so it is a reasonable approximation.}
in the non-local form factors. It is clear that
those are parts of the same logarithm because the source of both
is in the \  $\Tr\ln \big[ \cx + M^2 + \la \phi^2/2 \big]$. In the IR,
where $M^2$ term dominates, in Euclidean signature we meet
the expansion of the type
\beq
\ln \bigg( \frac{M^2 + k^2 + \la \phi^2/2}{\mu^2} \bigg)
\,\,\approx\,\,
\ln \Big( \frac{M^2}{\mu^2} \Big)
\,+\,
\frac{k^2 + \la \phi^2/2}{M^2} \,+\, ...\,\,,
\label{IR-nolog}
\eeq
which means there are no physically relevant logarithms in the
IR, in both mentioned cases, i.e., for the nonlocal form factors
(or polarization operator), and for the effective potential.

The situation for the two-scalar case with a strong hierarchy of
masses $m \ll M$ is technically more complicated, but we can
apply the method described above to reduce the calculation to
the two single-scalar models.

After double diagonalization and the standard calculations for a
single scalar (described in detail in \cite{CorPot,OUP}), we arrive
at the renormalized effective potential
\beq
V^{(1)}_{ren}
\,\,=\,\,
\frac{\Pi^2_1(\phi,\chi)}{4(4\pi)^2}\,
\ln\bigg[\frac{\Pi_1(\phi,\chi)}{\mu^2}\bigg]
\,+\,
\frac{\Pi^ 2_2(\phi,\chi)}{4(4\pi)^2}\,
\ln\bigg[\frac{\Pi_2(\phi,\chi)}{\mu^2}\bigg]\,.
\label{eq58}
\eeq

Let us evaluate the regime where the conditions
\beq
M^2 \gg \phi^2,
\qquad
M^2 \gg m^2
\label{ourIR}
\eeq
hold and also assume $m^2 \to 0$. According to what
was discussed above, this situation can be regarded as an IR limit
for the effective potential. Our experience with the decoupling
implies that the contribution of the large-mass field should vanish
at least quadratically and, in the first place, there should not be
relevant logarithms with $M$. To see whether this is the case, one
has to evaluate how the expressions for both $\Pi_j$ behave in this
regime.

It is easy to verify that the coefficient defined in (\ref{eq7}) has
the limit $\th \to \infty$. Then, the coefficients $a$, $b$ and $c$
in (\ref{eq10}) become
\beq
&&
a = \left( \frac12\, + \frac{\th}{2\sqrt{1 + \th^2}} \right) \approx 1,
\nn
\\
&&
b = \left( \frac12\, - \frac{\th}{2\sqrt{1 + \th^2}} \right) \approx 0,
\nn
\\
&&
c \,=\, \frac{1}{\sqrt{1 + \th^2}} \,\,\approx \,\,
\frac{2\la_{12}}{M^2}\,\phi\chi \,\,\approx \,\, 0\,.
\label{abc-IR}
\eeq
These relations mean that the mixing of two \textit{quantum}
scalars disappears, exactly like in the $\la_{12} \to 0$ case with
unrestricted masses $m$ and $M$. In this approximation, the
expressions for $\Pi_{1,2}$ reduce to (we intend to take $m \to 0$
limit at the end)
\beq
&&
\Pi_1 \,\approx \, M_{11}^2 \,\approx \,
-\,m^2 - \frac{\la_1}{2}\phi^2 - \frac{\la_{12}}{4}\chi^2.
\nn
\\
&&
\Pi_2 \,\approx\, M_{22}^2 \,-\, \frac{2\la_{12}}{M^2}
\,\phi\chi\,M_{12}^2
\, \approx \,-\,M^2.
\label{Pi12-IR}
\eeq

Let us skip the routine calculations and just go to the final result.
In the effective potential, the dominant contribution with
``surviving'' logarithm has  the form
\beq
&&
V^{(1)}
\,\,=\,\,
\frac{1}{4(4\pi)^2}\,\Big(\,m^2 + \frac{\la_1}{2}\phi^2
+ \frac{\la_{12}}{4}\chi^2\Big)^2\,
\ln\bigg(\frac{m^2 + \frac{\la_1}{2}\phi^2
+ \frac{\la_{12}}{4}\chi^2}{\mu^2}\bigg)
\nn
\\
&&
\qquad
\qquad
+\,\,
\frac{ M^4}{4(4\pi)^2}\,
\ln\bigg(\frac{ M^2}{\mu^2}\bigg) \,.
\label{almost final logs}
\eeq
The last term is the usual contribution to the cosmological constant,
which is of no interest to us. Omitting this term and taking the limit
$m \to 0$, the expression  (\ref{almost final logs}) becomes
\beq
&&
V^{(1)}\,=\,
\frac{\la^2 \rho^4}{16\,(4\pi)^2}\,
\ln\Big(\frac{\la \rho^2}{2\,\mu^2}\Big),
\quad
\mbox{where}
\quad
\frac{\la}{2}\, \rho^2
\,=\,
\frac{\la_1}{2}\phi^2 + \frac{\la_{12}}{4}\chi^2\,.
\label{final logs}
\eeq
Thus, the effective potential boils down to the leading logarithmic
contribution of a single scalar field $\rho$ that is a mixing of the
original scalars. When the original theory has no mixing,
\textit{i.e.},
when $\la_{12}=0$, this scalar coincides with the original light
scalar $\phi$. In any case, in the IR limit, as we defined it, only
the light scalar retains logarithmic contribution, while for
the large-mass scalar, there is a quadratic decoupling.

It remains to consider the IR decoupling in terms of
one-loop derivative expansions, that is $Z^{(1)}$'s. First of
all, the arguments about vanishing mixing (\ref{abc-IR}) in
the IR limit (\ref{ourIR}) remain valid, and hence it is
sufficient to consider only the single-scalar case (\ref{eq73}).
Taking the corresponding limit $M \gg \phi$ for the field with
the mass $M$, we get
\beq
Z^{(1)}_{IR}
\,\,=\,\,
\frac{\la^2}{12\,(4\pi)^2}\,
\bigg( 1 \,+\, \frac{\tilde{\xi}R}{2 M^2} \bigg)
\,\frac{\phi^2}{M^2}.
\label{eq73-IR}
\eeq
This formula shows that there is a quadratic decoupling and, for
$|R| \ll M^2$, the $R$-dependent term is vanishing with the double
(quartic) speed.

\section{Conclusions}
\label{Conc}

In this work, we developed a systematic approach to calculate
the effective potential and the first-order terms in the derivative
expansion in the theory with several scalar fields with different
masses. The \textit{double diagonalization} method is based
upon the functional approach to effective action
\cite{Jackiw74,iliopoulos1975,fraser1985} and consists from
\
\textit{i)} the diagonalization of the operator $\hat{H}$ (bilinear
form of the action) and \ \textit{ii)} changing the basis of
expansion to the linear independent directional derivatives in the
background scalar fields. The two-scalars calculation illustrating
this method, can be easily generalized to a larger number of
scalar fields. The use of this method enables a simple derivative
expansion not only in Minkowski spacetime, but also in curved
space, at least in the first order in scalar curvature.

The analysis of the two-scalar system with a strong hierarchy
of masses, when one mass dominates over the magnitudes of
scalars while the second mass remains small or even vanishes, has
shown the quadratic decoupling of quantum corrections in such an
IR limit. The contribution of the large-mass scalar reduces to the
one to cosmological constant term. This output is natural if we
remember the uniqueness of effective action and the known results
for the IR decoupling in the kinetic sector of a scalar field
\cite{Peixoto2003}.

From the viewpoint of applications, our results mean that the
quantum effects of two scalar fields in the assisted models of
inflation (see, e.g., \cite{Liddle_1998,Wands,Peterson:2010np})
boil down to the ones of the single lightest scalar in the course
of a fast expansion of the Universe, when the typical energy
becomes smaller, and the large-mass field decouples.

Let us discuss the possible extensions of the results of this work.
It is technically possible to construct the derivative extension
in the effective action of scalars owing to the fermion loops
\cite{Fraser-F}. Using the method presented above, one can
extend the existing expansions by including the curvature terms,
similar to Eq.~(\ref{eq73}).
The same concerns also the contributions of
gauge fields. In both cases the kinetic scalar term is UV divergent
and the coefficient $Z$ in (\ref{EffPot}) is subject of
renormalization procedure, however the power counting tells us
that this does not concern curvature-dependent part. We hope to
report on this calculation in future work.
In view of the possible applications to inflation, it looks
interesting to include also graviton  and mixed graviton-scalar
loops. However, this is a more complicated problem, because
quantum gravity based on GR is not renormalizable and because
such graviton contributions have high degree of ambiguity
(see the discussion and further references, e.g., in \cite{OUP}).
Anyway, this is an attractive idea which may be put into
practise in the effective quantum gravity framework.

\section*{Acknowledgements}

Authors are grateful to Oliver Piattella for explanations
concerning the inflationary models with two scalars.
A.G.B. is grateful to Universidade
Federal de Juiz de Fora (UFJF) for supporting her Ms.C. project.
I.Sh. gratefully acknowledges partial support from CNPq (Conselho
Nacional de Desenvolvimento Cient\'{i}fico e Tecnol\'{o}gico, Brazil)
under the grant 305122/2023-1.



\begin{thebibliography}{bib}

\bibitem{Liddle_1998}
A. R. Liddle, A. Mazumdar and F.E. Schunck,
{\it Assisted inflation,}
Phys. Rev. D {\bf 58} (1998) 061301.

\bibitem{Wands} D. Wands,
{\it Multiple field inflation,}
in {\it Inflationary Cosmology} (Springer Berlin, Heidelberg, 2007, 
Editors: Martin Lemoine, Jerome Martin, Patrick Peter)
pp. 275--304.

\bibitem{Peterson:2010np} C. M. Peterson and M. Tegmark,
{\it Testing Two-Field Inflation,}
Phys. Rev. D {\bf 83} (2011) 023522.

\bibitem{Heatal} M.~He, A.A.~Starobinsky and J.~Yokoyama,
\textit{Inflation in the mixed Higgs-$R^2$ model,}
JCAP \textbf{05} (2018) 064,
arXiv:1804.00409. 

\bibitem{iliopoulos1975}
J. Iliopoulos, C. Itzykson and A. Martin,
{\it Functional methods and perturbation theory,}
Rev. Mod. Phys. {\bf 47} (1975) 165. 

\bibitem{fraser1985} C.M. Fraser,
{\it Calculation of higher derivative terms in the one-loop
effective Lagrangian,}
Z. Phys. C {\bf 28} (1985) 101. 

\bibitem{Wein-II} S. Weinberg,
{\it The quantum theory of fields, 
Vol II.  Modern applications}  
(Cambridge University Press, Cambridge, 1996).

\bibitem{Manohar-2000} A.V.~Manohar and M.B.~Wise,
\textit{Heavy quark physics,}
(Cambridge Monogr. Part. Phys. Nucl. Phys. Cosmol. \textbf{10} 2000)

\bibitem{Hu-1984}
B.L.~Hu and D.J.~O'Connor,
\textit{Effective Lagrangian for $\lambda \phi^4$ theory in
curved space-time with varying background fields: quasilocal
approximation,}
Phys. Rev. \textbf{D30} (1984) 743.

\bibitem{Kirsten-1993}
K.~Kirsten, G.~Cognola and L.~Vanzo,
\textit{Effective Lagrangian for selfinteracting scalar field
theories in curved space-time,}
Phys. Rev. \textbf{D48} (1993) 2813, 
hep-th/9304092.

\bibitem{Cheyette-1985} O.~Cheyette,
\textit{Derivative Expansion of the Effective Action,}
Phys. Rev. Lett. \textbf{55} (1985) 2394.

\bibitem{Chan} L.-H. Chan,
\textit{Effective-Action expansion in perturbation theory,}
Phys. Rev. Lett. \textbf{54} (1985) 1222;
Erratum ibid.  \textbf{56} (1986) 404.

\bibitem{Bunch1979} T.S. Bunch and L. Parker,
{\it Feynman propagator in curved space-time: a momentum
space representation,}
Phys. Rev. D {\bf 20} (1979) 2499. 

\bibitem{OUP} I.L.~Buchbinder and I.L.~Shapiro,
\textit{ Introduction to quantum field theory with applications to
quantum gravity,} (Oxford University Press, Oxford, 2021).

\bibitem{Petrov} A.Z.~Petrov, {\it Einstein Spaces,}
(Pergamon Press, Oxford, 1969).

\bibitem{Tensors} I.L.~Shapiro,
{\it Primer in tensor analysis and relativity,}
(Springer, NY, 2019).

\bibitem{ColeWein} S.R. Coleman and E.J. Weinberg,
{\it Radiative corrections as the origin of spontaneous symmetry
breaking,} Phys. Rev. {\bf D7} (1973) 1888.

\bibitem{Jackiw74} R.~Jackiw,
\textit{Functional evaluation of the effective potential,}
Phys. Rev. \textbf{D9} (1974) 1686.

\bibitem{CorPot} F. Sobreira, B.J. Ribeiro and I.L. Shapiro,
{\it Effective potential in curved space and cut-off regularizations,}
Phys. Lett. B {\bf 705} (2011) 273, 
arXive: 1107.2262. 

\bibitem{Shore1979} G.M.~Shore,
\textit{Radiatively induced spontaneous symmetry breaking and
phase transitions in curved space-time,}
Annals Phys. \textbf{128} (1980) 376.

\bibitem{Buchbinder1985} I.L.~Buchbinder and S.D.~Odintsov,
\textit{Effective potential and phase transitions induced by
curvature in gauge theories in curved space-time,}
Yad. Fiz. \textbf{42} (1985) 1268. 
Class. Quantum Grav. \textbf{2} (1985) 721,

\bibitem{Buchbinder1988} I.L.~Buchbinder and J.J.~Wolfengaut,
\textit{Renormalization group equations and effective action in
curved space-time,}
Class. Quant. Grav. \textbf{5} (1988) 1127. 

\bibitem{Asorey2022}
M.~Asorey, W.C.~Silva, I.L.~Shapiro and P.R.B.d.~Vale,
\textit{Trace anomaly and induced action for a metric-scalar
background,}
Eur. Phys. J. \textbf{C83} (2023) 157.
arXiv:2202.00154.

\bibitem{Peixoto2003}
G.~de Berredo-Peixoto, E.V.~Gorbar and I.L.~Shapiro,
\textit{On the renormalization group for the interacting massive
scalar field theory in curved space,}
Class. Quant. Grav. \textbf{21} (2004) 2281, 
hep-th/0311229.

\bibitem{bavi90} A.O.~Barvinsky and G.A.~Vilkovisky,
{\it Covariant perturbation theory. 2: Second order in the curvature.
General algorithms,}
Nucl. Phys. {\bf 333B} (1990) 471.

\bibitem{apco} E.V.~Gorbar and I.L.~Shapiro,
{\it Renormalization Group and Decoupling in Curved Space},
JHEP {\bf 02} (2003) 021,
hep-ph/0210388.

\bibitem{Omar-FF4D}
S.A.~Franchino-Vi\~nas, T.~de Paula Netto, I.L.~Shapiro
and O.~Zanusso,
{\it Form factors and decoupling of matter fields in
four-dimensional gravity,}
Phys. Lett. {\bf B790} (2019) 229,
arXiv:1812.00460.

\bibitem{Barvinsky1990-34} A.O.~Barvinsky and G.A.~Vilkovisky,
\textit{Covariant perturbation theory. 3: Spectral representations
of the third order form-factors,}
Nucl. Phys. \textbf{B333} (1990) 512; 
 \\
A.O.~Barvinsky, Y.V.~Gusev, V.V.~Zhytnikov and G.A.~Vilkovisky,
\textit{Covariant perturbation theory. 4. Third order in the curvature,}
arXiv:0911.1168.

\bibitem{Buchbinder2019}
I.L. Buchbinder, A.R. Rodrigues, E.A. dos Reis and I.L. Shapiro,
{\it Quantum aspects of Yukawa model with scalar and axial scalar
fields in curved spacetime,}
Eur. Phys. J. C {\bf 79} (2019) 1002,
arXiv:1910.01731.

\bibitem{AC} T.~Appelquist and J.~Carazzone,
{\it Infrared Singularities and Massive Fields,}
Phys. Rev.  {\bf D11} (1975) 2856.

\bibitem{EH-1938} W. Heisenberg and H. Euler,
\textit{Folgerungen aus der Diracschen Theorie des Positrons},
Zeitschrift für Physik  \textbf{98} (1936) 
714. 

\bibitem{sponta} E.V.~Gorbar and I.L.~Shapiro,
\textit{ Renormalization group and decoupling in curved space.
3. The Case of spontaneous symmetry breaking,}
JHEP \textbf{02} (2004) 060,
hep-ph/0311190.

\bibitem{Tmn-sponta}
M.~Asorey, P.M.~Lavrov, B.J.~Ribeiro and I.L.~Shapiro,
\textit{ Vacuum stress-tensor in SSB theories,}
Phys. Rev. \textbf{D85} (2012) 104001;
arXiv:1202.4235.

\bibitem{Fraser-F} I.J.R.~Aitchison and C.M.~Fraser,
\textit{ Fermion Loop Contribution to Skyrmion Stability,}
Phys. Lett. \textbf{B146} (1984) 63;  
\textit{ Derivative expansions of fermion determinants: anomaly
induced vertices, Goldstone-Wilczek currents and Skyrme terms,}
Phys. Rev.  \textbf{D31} (1985) 2605.

\end{thebibliography}
\end{document}